\documentclass[aps,prl,twocolumn,groupedaddress,superscriptaddress,floatfix]{revtex4-1}
\usepackage{graphicx}
\usepackage{dcolumn}
\usepackage{bm}
\usepackage{amsmath}
\usepackage{amssymb}
\usepackage{hyperref}
\usepackage{color}

\makeatother

\begin{document}
\title{Nonlinear looped band structure of Bose-Einstein condensates in an optical lattice}

\author{S. B. Koller}
\affiliation{Joint Quantum Institute, National Institute of Standards and Technology and University of Maryland, Gaithersburg, Maryland 20899, USA} 
\affiliation{Physikalisch Technische Bundesanstalt, Braunschweig, Germany}

\author{E. A. Goldschmidt}
\affiliation{United States Army Research Laboratory, Adelphi, Maryland 20783, USA}

\author{R. C. Brown}
\altaffiliation{Current address: National Institute of Standards and Technology, Boulder, Colorado 80305 USA}
\affiliation{Joint Quantum Institute, National Institute of Standards and Technology and University of Maryland, Gaithersburg, Maryland 20899, USA}

\author{R. Wyllie}
\altaffiliation{Current address: Quantum Systems Division, Georgia Tech Research Institute, Atlanta, Georgia 30332, USA}
\affiliation{Joint Quantum Institute, National Institute of Standards and Technology and University of Maryland, Gaithersburg, Maryland 20899, USA}

\author{R. M. Wilson}
\affiliation{Department of Physics, United States Naval Academy, Annapolis, Maryland 21402, USA}

\author{J. V. Porto}
\affiliation{Joint Quantum Institute, National Institute of Standards and Technology and University of Maryland, Gaithersburg, Maryland 20899, USA}

\begin{abstract}
We study experimentally the stability of excited, interacting states of bosons in a double-well optical lattice in regimes where the nonlinear interactions are expected to induce ``swallowtail'' looped band structure. By carefully preparing different initial coherent states and observing their subsequent decay, we observe distinct decay rates that provide direct evidence for multivalued, looped band structure. The double well lattice both stabilizes the looped band structure and allows for dynamic preparation of different initial states, including states within the loop structure. We confirm our state preparation procedure with dynamic Gross-Pitaevskii calculations. The excited loop states are found to be more stable than dynamically unstable ground states, but decay faster than expected based on a mean-field stability calculation, indicating the importance of correlations beyond a mean field description.
\end{abstract}

\date{\today}
\maketitle
Interactions in Bose-Einstein condensates (BECs) can give rise to qualitatively new nonlinear phenomena \cite{Denschlag07012000,PhysRevLett.101.130401,PhysRevLett.83.5198,NatPhys.7.1.61,fourwavemixing}. For example, superfluids in optical lattices can exhibit additional, interaction-stabilized states arising from the so-called ``swallowtail catastrophe" in which the band structure becomes multi-valued \cite{NJP1367-2630-5-1-104,PhysRevA.66.013604,PhysRevA.72.033602,PhysRevA.67.053613}.
As the interaction increases, the collective band structure at the edge of the Brilloiun zone (BZ) develops a cusp (a discontinuity in the derivative), and subsequently a loop with multiple energy states that can be occupied at the same quasimomentum. The existence of loop states is related to dynamical asymmetry in Landau-Zener tunneling between coupled states of the many-body system \cite{PhysRevA.61.023402}, which has been used to indirectly observe nonlinear loop structure \cite{PhysRevLett.91.230406,NatPhys.7.1.61}. Despite the fact that ultracold atoms in optical lattices are an ideal system to realize nonlinear wave dynamics, the interaction strengths needed to generate such interesting band structure in a simple lattice are prohibitively large. 

\begin{figure}
  \centering
  \includegraphics[width=0.95\linewidth]{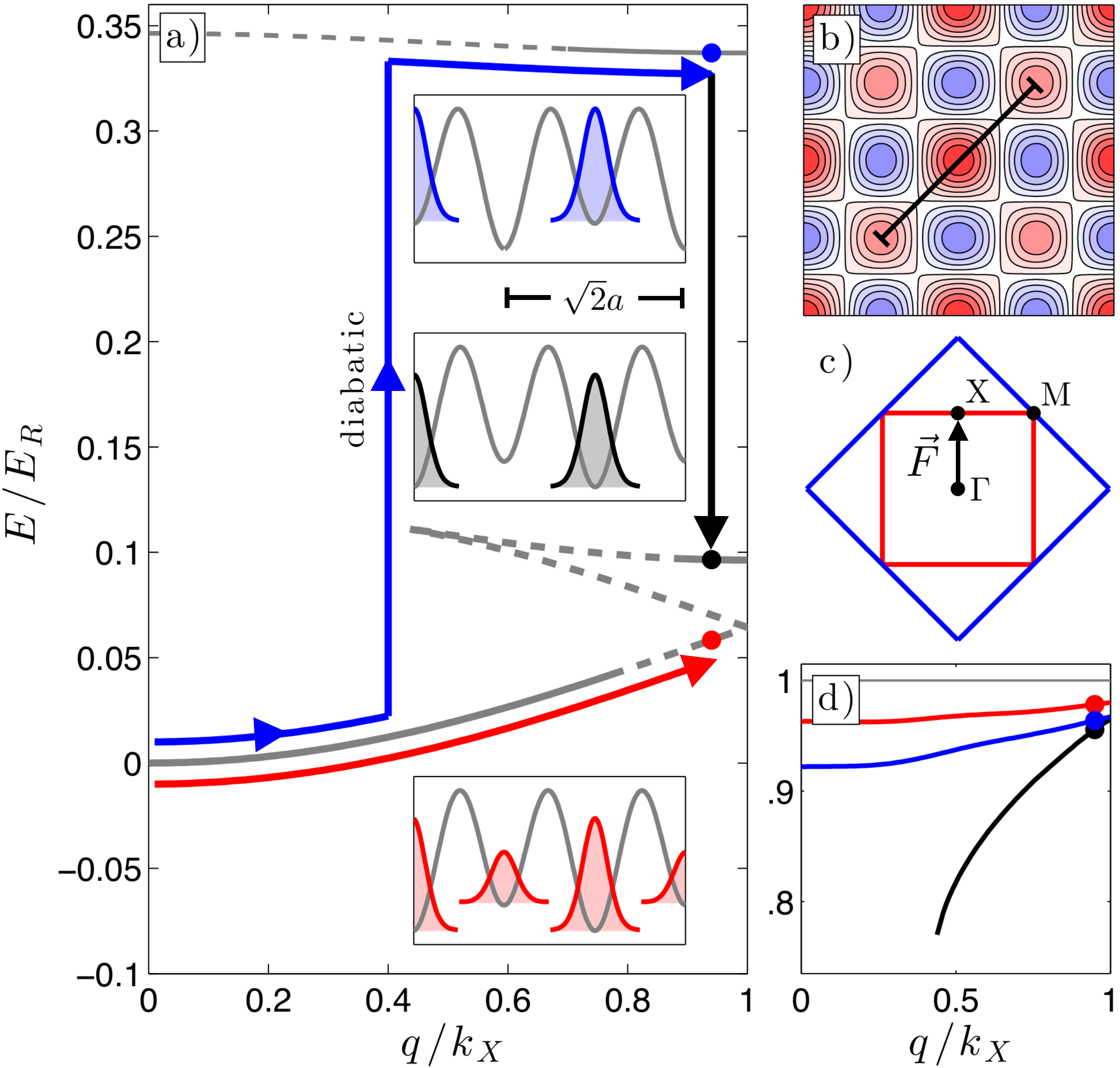}
  \caption{\label{fig:prep} (a) State preparation of an interacting BEC in the looped band structure (grey solid and dashed lines). Dashed parts of the band structure indicate dynamically unstable regions. A combination of forces to accelerate the quasimomentum and control of the staggered offset $\Delta$ are used to prepare the BEC in the ground (red), loop (black), or excited (blue) bands as indicated schematically by the lines with arrows and described in the text. Note that until the final state preparation step, the staggered offset is too large to support a loop. The insets show the final wavefunctions calculated by a time-dependent Gross-Pitaevskii (GP) simulation. (b) Real-space lattice potential with staggered wells in 2D. (c) 2D Brillouin zones associated with the lattice of period $\lambda/2$ (blue) and $\lambda/\sqrt{2}$ (red). Acceleration is from $q=0$ at $\Gamma$ to $q=k_X$ at X. (d) Overlap between the dynamic GP simulation and the ideal wavefunctions for the E, L, and G preparations sequences, as a function of $q$. 
}
\end{figure}

In addition to multi-valued band structure at the edge of the BZ, period doubled solutions are also expected to occur halfway to the edge of the BZ~\cite{PhysRevA.69.043604}. Adding a weak lattice at half the main lattice period expands the parameter regime where band structure loops are expected \cite{PhysRevA.72.033602}, making them more experimentally feasable. The states associated with the loop are collective excited states, and an essential consideration in their observation is their stability. Even in the weakly interacting, mean-field limit, dynamical instabilities \cite{PhysRevA.77.012712,PhysRevLett.86.4447,PhysRevA.64.061603} can arise that quickly destroy the excited superfluid state. Dynamically stable mean-field solutions exist 
\cite{PhysRevA.86.063636}, and in particular there are accessible regimes where mean field calculations predict different stability for the multi-valued bands. An example of such looped band structure is shown in Fig.~\ref{fig:prep}a. Correlations outside of a mean field description of the system, however, can cause additional instability in the excited states~\cite{NatPhys.7.1.61,altman05a}. Using ultra cold atoms to study unconventional excited states~\cite{stojanovic08a, wu09b, olschlager11a, Wirth2011a, Kock2015} requires understanding such relaxation processes.

Here, we dynamically produce nonlinear excited states of a BEC in a two dimensional optical lattice and show experimentally that the multi-valued nature of the looped band structure can be observed as differences in the stability of BEC coherence, depending on which initial nonlinear state is prepared. The distinct coherence decay rates occur near the band edge for addition of a weak lattice at half the main period, in qualitative agreement with theory. We observe substantial decay in the loop states where a mean-field treatment predicts stability, indicating that inhomogeneities and correlations that invalidate the mean-field description may play an important role~\cite{altman05a}. By measuring the energy released upon decay, we show that there is an energy difference between states prepared in the loop and ground band, providing additional evidence of multi-valued band structure. 

Although equilibrium calculations suggest the existence of stable loop states at the band edge, such states are not necessarily trivial to produce experimentally. The mean-field interacting states obey Bloch's equation and are characterized by a quasimomentum $q$. Raman or Bragg excitation can excite weakly-interacting BECs to a given quasimomentum $q\neq 0$ because the initial and final states are single-particle in nature, as are such excitation techniques. The loop states, however, rely on interactions, and it is not clear how well single-particle Raman excitation couples to such collective states via intermediate states with only partial transfer.  In order to prepare states in the interacting band with high fidelity, we use a combination of adiabatic and diabatic manipulation of the lattice structure~\cite{Wirth2011a, Anderlini2007} while accelerating the BEC to momenta near the band edge. 

The 2D lattice is produced with a 813~nm laser in a bow-tie configuration and weak harmonic confinement in the third dimension, giving rise to a checkerboard array of 1D tubes~\cite{PhysRevA.73.033605}. The staggered energy offset, $\Delta$, between neighboring tubes can be dynamically controlled on timescales as short as 10~$\mu$s, faster than any dynamic timescale in the system. For $\Delta \neq 0$, the fundamental lattice period increases from the usual $\lambda/2$ to $ \lambda/\sqrt{2}$, and the Bravais lattice and associated BZ are rotated 45 degrees with respect to the original $\lambda/2$ lattice (Fig.~\ref{fig:prep}c). A stable looped structure is expected along the edge of the smaller BZ for $\Delta$ on the order of interaction energy parameterized by $g\bar{n}$ ($\bar{n}$ is the average atomic density~\cite{supplement} and $g=4\pi\hbar^2a/m$, where $a$ is the scattering length and $m$ the mass of $^{87}$Rb). We develop three different procedures (described below), to prepare initial states that we label ``(G)round", ``(E)xcited'' and ``(L)oop" and then study the stability of each of these states by measuring characteristic decay timescales as functions of quasimomentum $q$, staggered offset $\Delta$, and atomic density $n$. 

All experiments start with a $^{87}$Rb BEC at rest ($q=0$) in the ground band of a $10.6~E_R$ lattice with a staggered offset too large to support a loop. (Here, $E_R=\hbar k^2_R /2m=3.5~\rm{kHz}$ is the single-photon recoil energy associated with the short period lattice, $k_R = 2 \pi/\lambda=\sqrt{2}k_X$.) A force, $F$, is applied to the atoms using a magnetic field gradient, resulting in an acceleration $\dot{q}=F$ along the direction ``X" associated with the long period of the lattice (see Fig.~\ref{fig:prep}) to near the edge of the band, which occurs at $q=k_X=\pi/(\lambda/\sqrt{2})$. The force is chosen to accelerate fast enough to minimize the decay associated with dynamic instabilities \cite{PhysRevA.64.061603}, yet slow enough to preclude band excitation. The resulting acceleration rate is calibrated by pulling the atoms through the entire BZ and observing Bloch oscillations of $q$ \cite{PhysRevLett.87.140402}. 

For each state preparation sequence, G, L, and E, the final lattice configuration is identical, with a final staggered offset $\Delta$ of order the interaction energy $g\bar{n}$. For the G sequence, the accelerating force is applied until the desired final $q$ is reached, at which point the offset is reduced (in 50~$\mu$s) to the final value $\Delta$. For the L and E sequences, when the accelerating force brings the BEC to $q\approx k_X/2$, the sign of the staggered offset is switched (in 50~$\mu$s), projecting the state into the excited band. Switching at $q\approx k_X/2$ avoids the dynamically unstable regions $0<q<k_X/2$ in the excited band and $k_X/2<q<k_X$ in the ground band. The accelerating force continues to increase $q$ with the BEC in the excited band until the desired $q$ is reached, at which time the offset is switched to the final $\Delta$ and the accelerating force is terminated. The wavefunctions in the excited and loop bands are nearly identical for $\Delta \rightarrow -\Delta$ when $q$ is near $k_X$ so the sign of the final $\Delta$ determines whether the E or L state is prepared. See \cite{supplement} for further information on the state preparation. 

\begin{figure}
  \centering
  \includegraphics[width=0.992\linewidth]{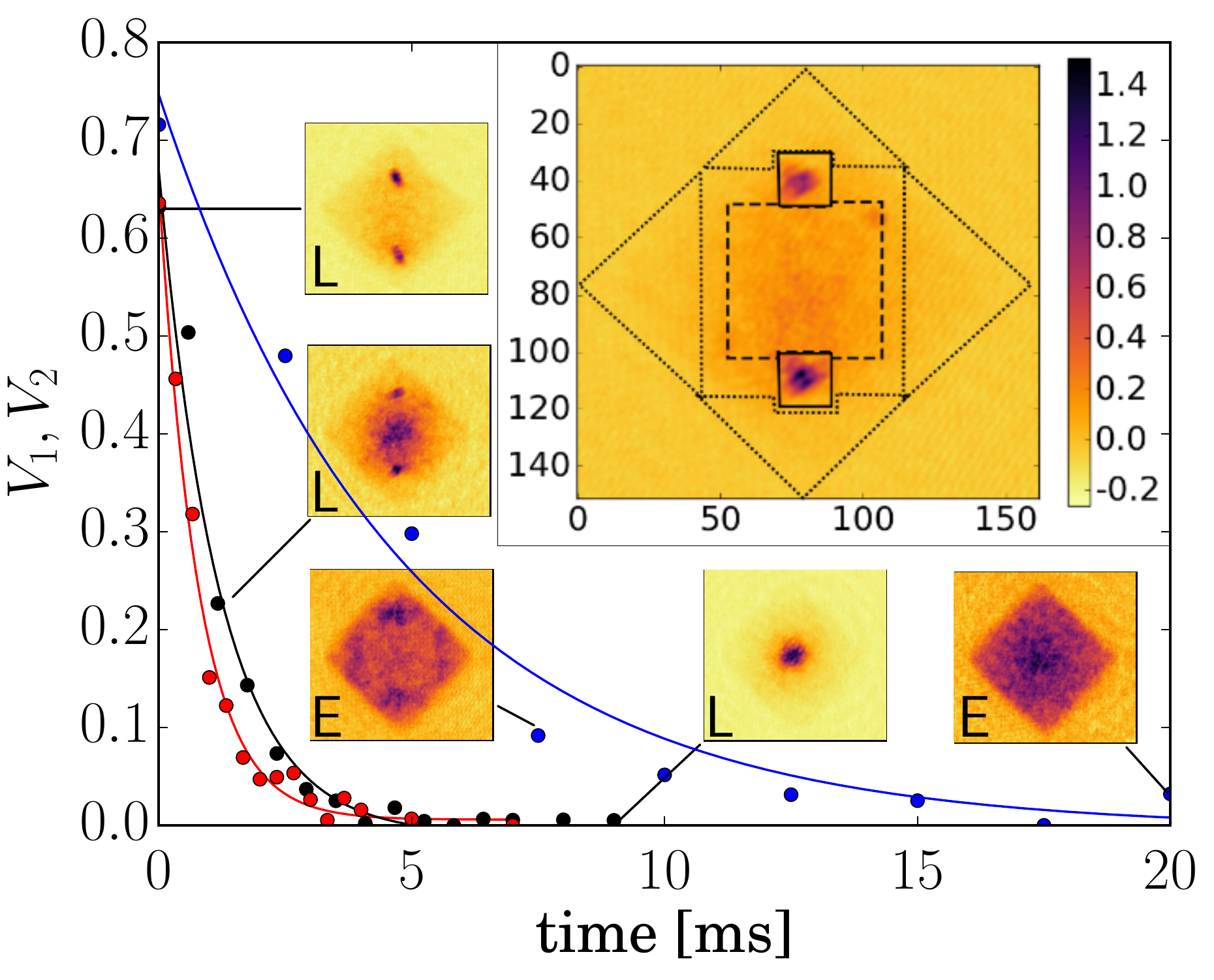}
  \caption{\label{fig:decay} Decay of visibility as a function of hold time for states prepared by the Ground (red), Loop (black), Excited (blue) sequence. The examples here correspond to the case of $q=k_X$, $\Delta/h$=0.7~kHz and $g\bar{n}/h$=0.31(1)~kHz. The fitted rates for the G-, L-, and E-preparations are $1.29(9)~\textrm{ms}^{-1}$, $0.84(8)~\textrm{ms}^{-1}$, and $0.21(3)~\textrm{ms}^{-1}$, respectively. Large inset: absorption image with analysis regions $S$ (solid lined squares), $B_1$ (dashed line central square), and $B_2$ (dotted lined triangles)~\cite{supplement}. Smaller insets: absorption images at different decay times, as indicated. 
  }
\end{figure}

To confirm the preparation sequences, we simulated the dynamics using the time-dependent Gross-Pitaevskii (GP) equation with the time sequences used in the experiment, starting with the GP ground state at $q=0$. We then calculated the overlap of the dynamically created nonlinear wavefunctions with the ideal GP solutions at the same $q$ and $\Delta$. The resulting overlap with the given target state is better than 90$~\%$ for the range of $q$ studied for all three sequences, as shown by the solid lines in Fig.~\ref{fig:prep}b. Note that unlike for linear equations, two different nonlinear solutions to the same GP equation are not expected to be orthogonal, and overlap with the unwanted states does not imply preparation infidelity. Indeed, there is significant overlap for the ideal GP solutions between the different solutions. We note that non-adiabatic excitation of the states can contribute to the instability of dynamically prepared states.

Having prepared states near the desired state in a given lattice configuration, the BEC is held in the lattice for a variable time $t_D$. The lattice is subsequently turned off in 1.5~ms, chosen such that the quasimomentum distribution is mapped to position following time-of-flight~\cite{supplement}. We take an absorption image and determine the occupation of the first and second BZ for each preparation sequence and $t_D$. To quantify the decay, the absorption signal is integrated over regions of the BZ that contain the initial coherent BEC to get the average column density $S$ within those small regions. The signal $S$ is compared to the integrated column density $B_1$ ($B_2$) contained in the first (second) BZ, with visibility defined as \cite{supplement}  
\begin{equation}
  V_i=\frac{S-B_i}{{S+B_i}} + C,
\end{equation}
where $C$ is chosen so that the visibility $V_i$ decays to zero. Example  $V_i$ are shown in Fig.~\ref{fig:decay}, with example images of the filling of the first and second BZ as insets. The large inset indicates the regions  $S$, $B_1$, and $B_2$.

The behavior of the decay is qualitatively and quantitatively different for the different preparation sequences: the E sequence leads to slower decay (near the band edge) that initially fills the second BZ before eventually filling the first BZ, while the G and L sequences lead to decay that fills predominantly the first BZ. As such, $V_1$ is used to extract decay rates for the G and L preparations and $V_2$ is used to extract a decay rate for the E prepared states. Despite the fact that the final lattice configurations are identical for the G and L sequence, we find that the L prepared states decay more slowly than the G states over a range of parameters.

\begin{figure}
  \centering
    \includegraphics[width=0.492\linewidth]{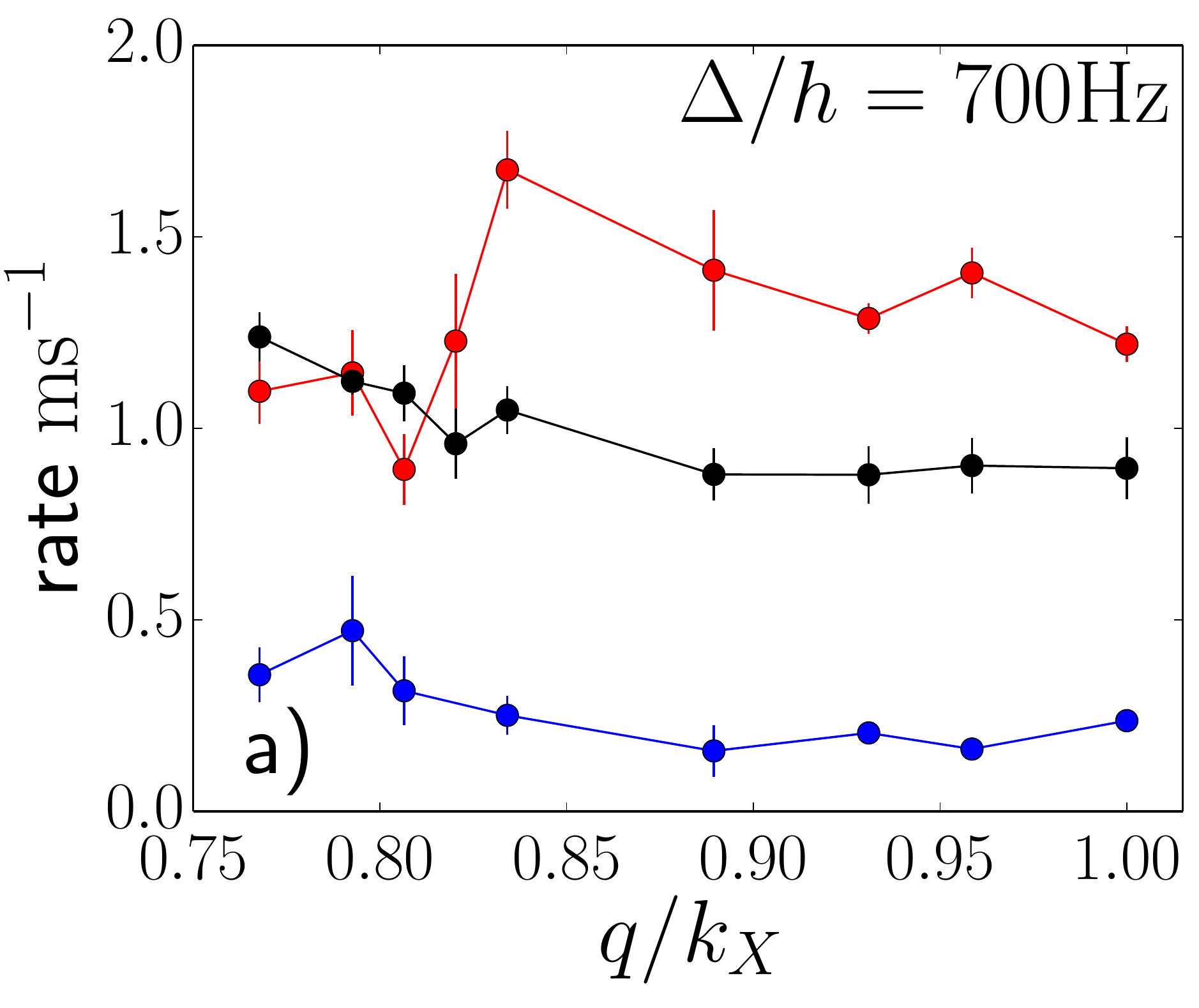}
    \includegraphics[width=0.492\linewidth]{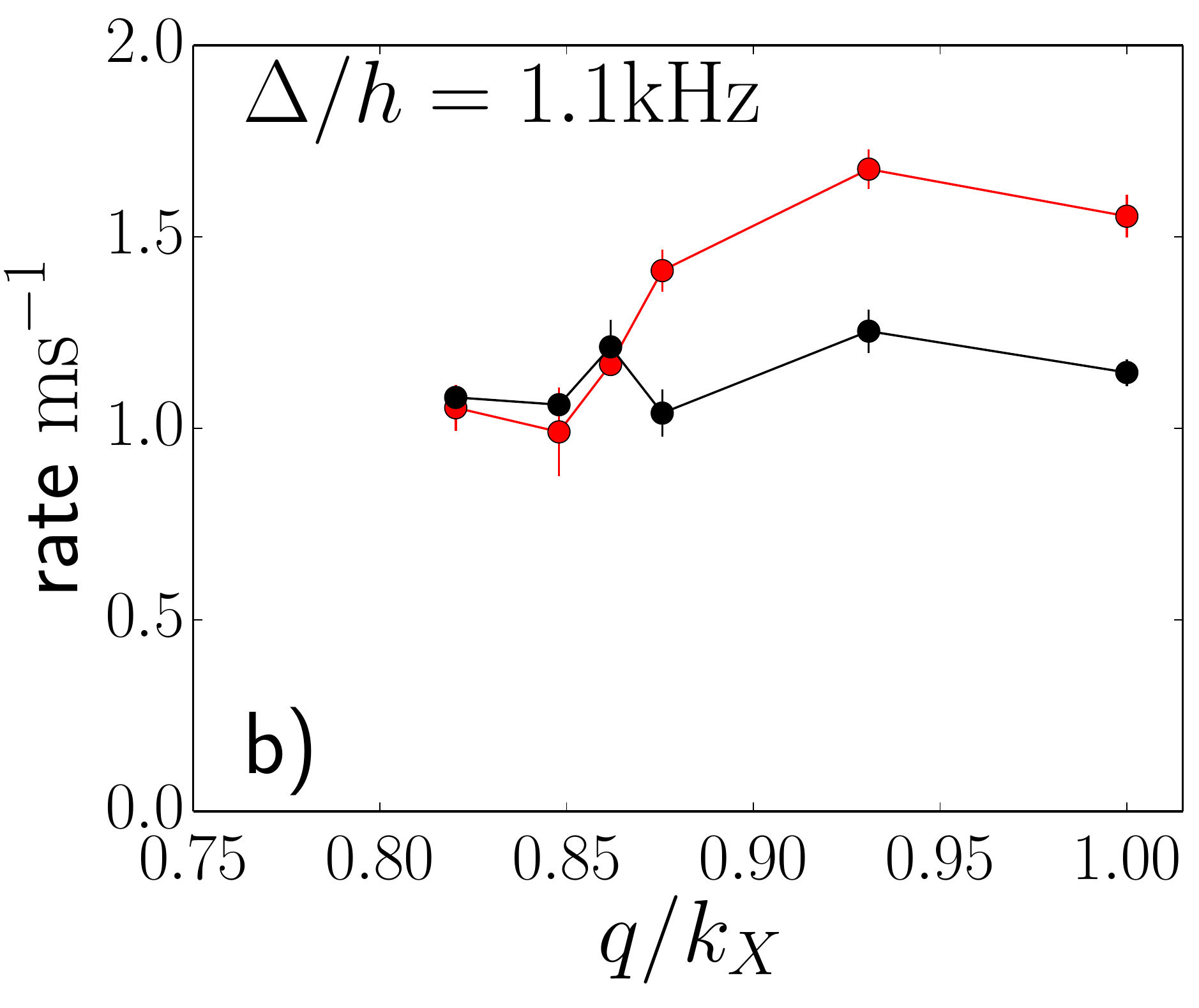}
    \caption{\label{fig:qscans} Measured decay rates vs quasimomentum $q$ of the G (red), L (black), and E (blue) states for a) $\Delta/h$=0.7~kHz and and $g\bar{n}/h$=0.31(1)~kHz and b) $\Delta/h$=1.1~kHz and $g\bar{n}/h$=0.29(1)~kHz. The color scheme for the G (black), L (red) and E (blue) prepared states is the same as in Fig.~\ref{fig:decay} 
}
\end{figure}

We first study the stability of the G, L, and E states as a function of quasimomentum $q$. Fig. \ref{fig:qscans} shows the decay rates for the ground and loop states for two different staggered offsets $\Delta/h=0.7~\rm{kHz}, 1.1~\rm{kHz}$ and similar interaction energy $g\bar{n}/h\approx0.31~\rm{kHz}$. The slower, excited band decay (in blue) is shown primarily for reference when included.
We find the L decay rate is approximately $40~\%$ smaller than the G decay rate for $q$ near the band edge, indicating greater stability of the loop states in that region. The discrepancy in decay rate disappears, however, at smaller $q$. The closing of the discrepency between the L and G decay rates occurs at larger $q$ for larger staggered offset $\Delta$, qualitatively agreeing with GP calculations \cite{PhysRevA.86.063636}. Near $q=k_X$, a GP analysis predicts dynamical stability, which we do not observe~\cite{supplement}. The relatively large decay rates of the L state as compared to the higher energy E state may be due to inhomogeneities in the system that close the loop when the local density becomes too low to support a loop, or due to correlations~\cite{PhysRevA.79.042703} of the atoms in the lattice, invalidating the mean-field description in \cite{PhysRevA.86.063636}.

\begin{figure}
\centering
  \includegraphics[width=0.492\linewidth]{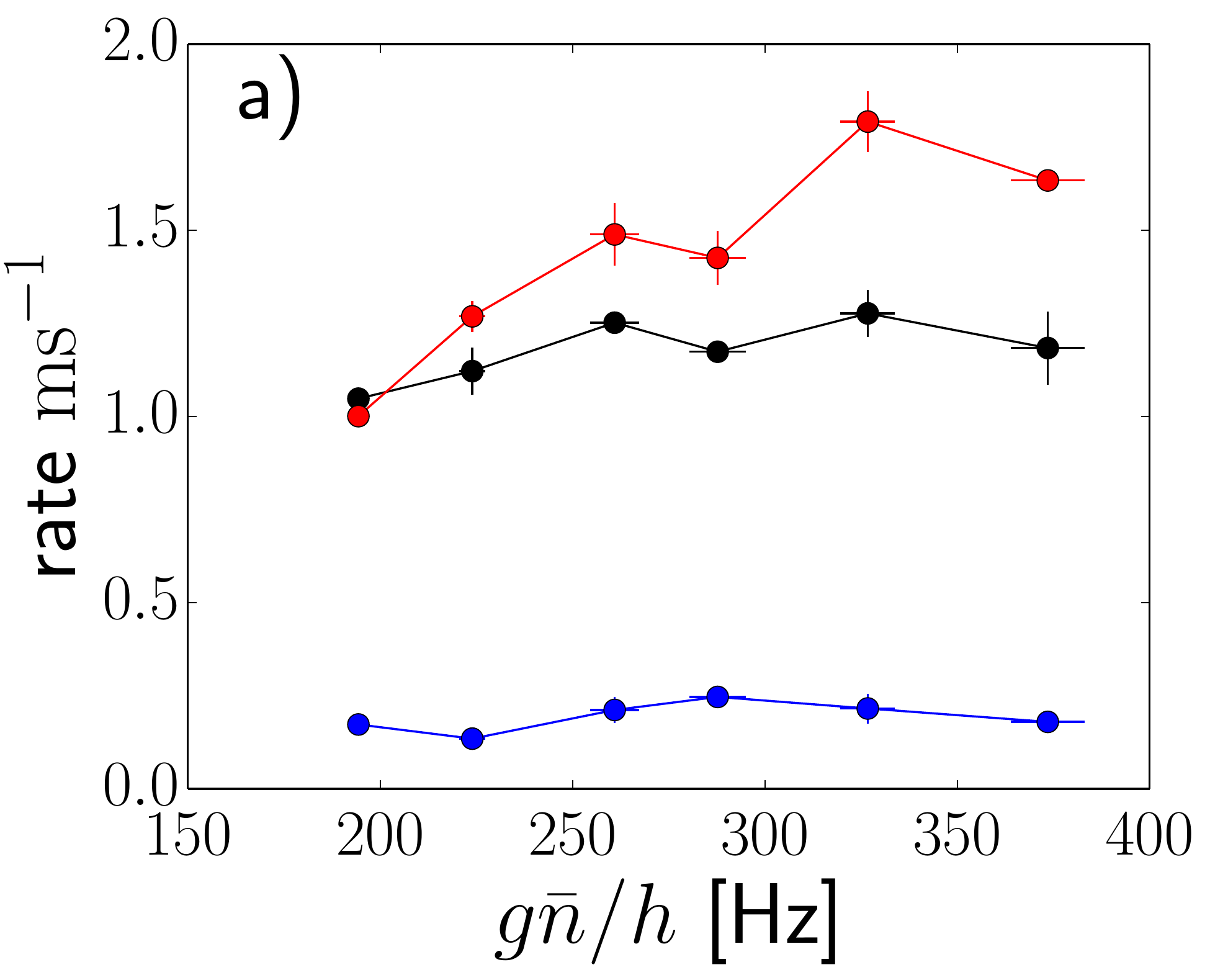}
  \includegraphics[width=0.492\linewidth]{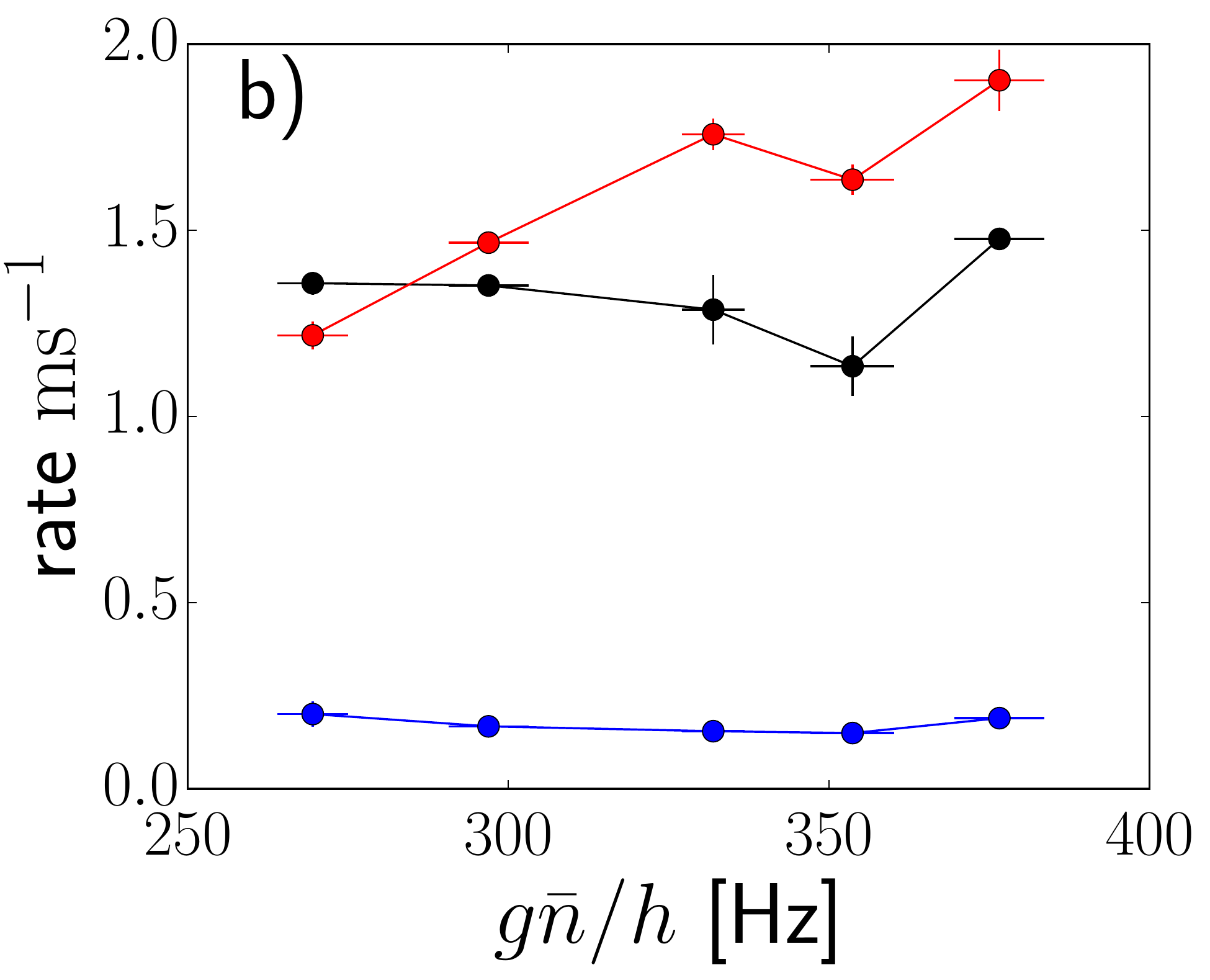}\\
  \includegraphics[width=0.492\linewidth]{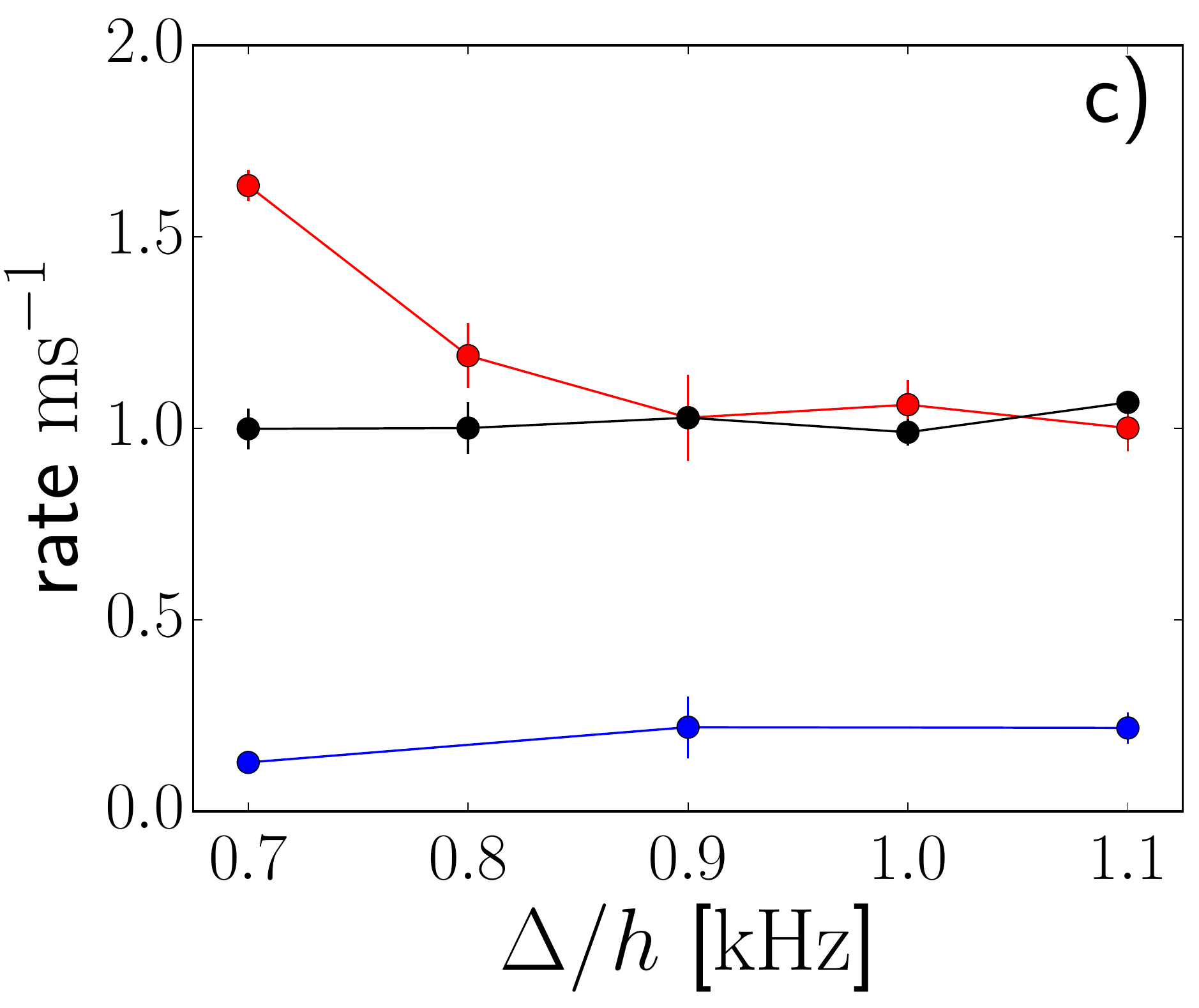}
  \includegraphics[width=0.492\linewidth]{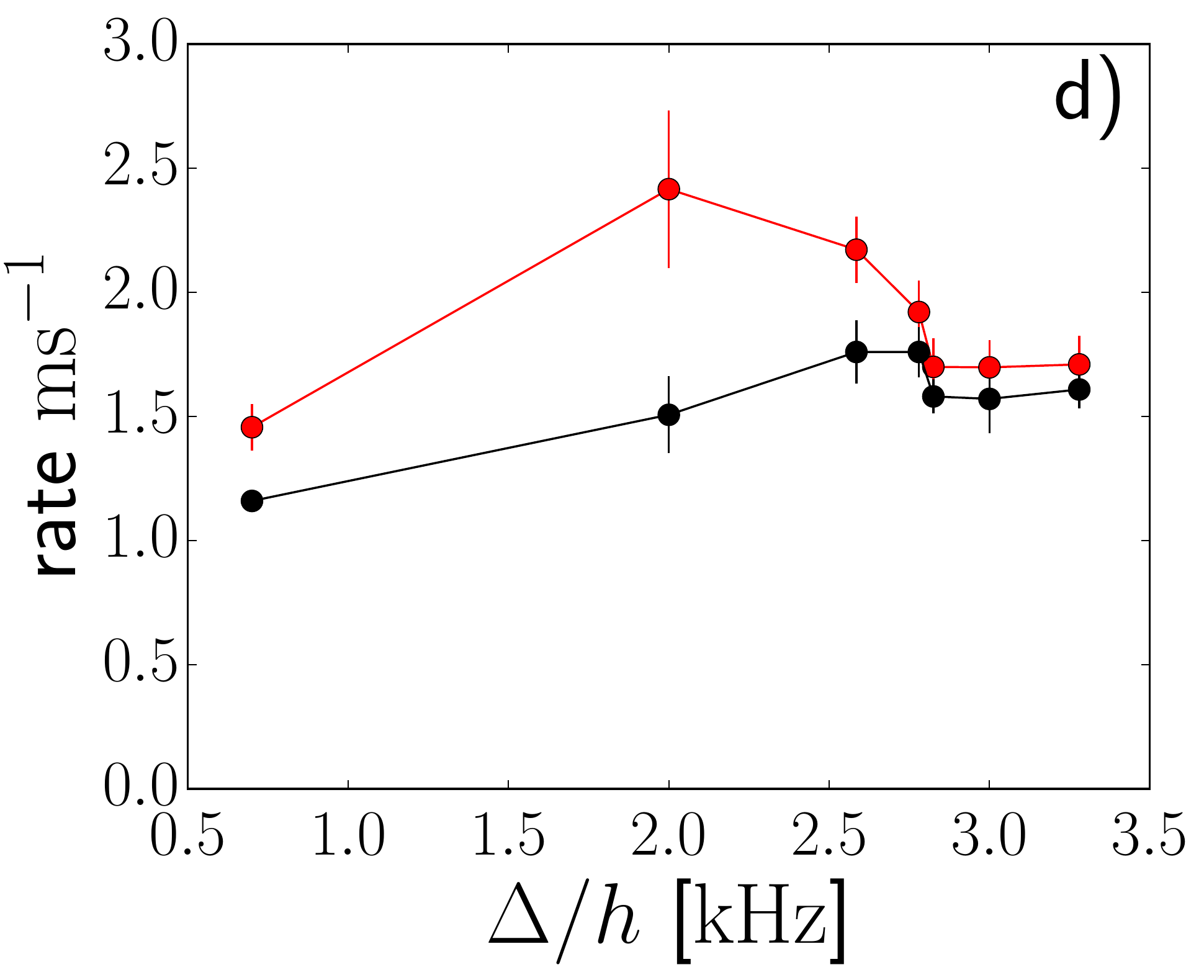}
  \caption{\label{fig:Dscans} Decay rates vs interaction energy $g\bar{n}$ for different states with $q=k_X$ and a) $\Delta/h$=0.9~kHz and b) $\Delta/h$=1.1~kHz. Decay rates vs offset $\Delta/h$ with c) $q=0.82~k_X$ and $g\bar{n}/h=0.29(1)~\rm{kHz}$ and d) $q=k_X$ and $g\bar{n}/h=0.31(1)~\rm{kHz}$. Vertical error bars indicate the standard deviation from the fit of the decay and horizontal error bars on a) and b) indicate the standard deviation of the mean of all measured instances. 
  }
\end{figure}

We expect the loop to only be present for $\Delta$ smaller than the interaction energy. We observe this behavior, as the difference between the L and G decay rates decreases for increasing $\Delta$ or decreasing atomic density (see Fig. \ref{fig:Dscans}). In particular, the interaction necessary to observe a difference between the G and L state is higher for larger staggered offset, as seen in Fig. \ref{fig:Dscans}a-b. In addition, Fig. \ref{fig:Dscans}c-d shows that the loop and ground decay rates converge closer to the band edge for larger staggered offset, as in Fig.~\ref{fig:qscans}.

\begin{figure}[!t]
\centering
    \includegraphics[width=0.492\linewidth]{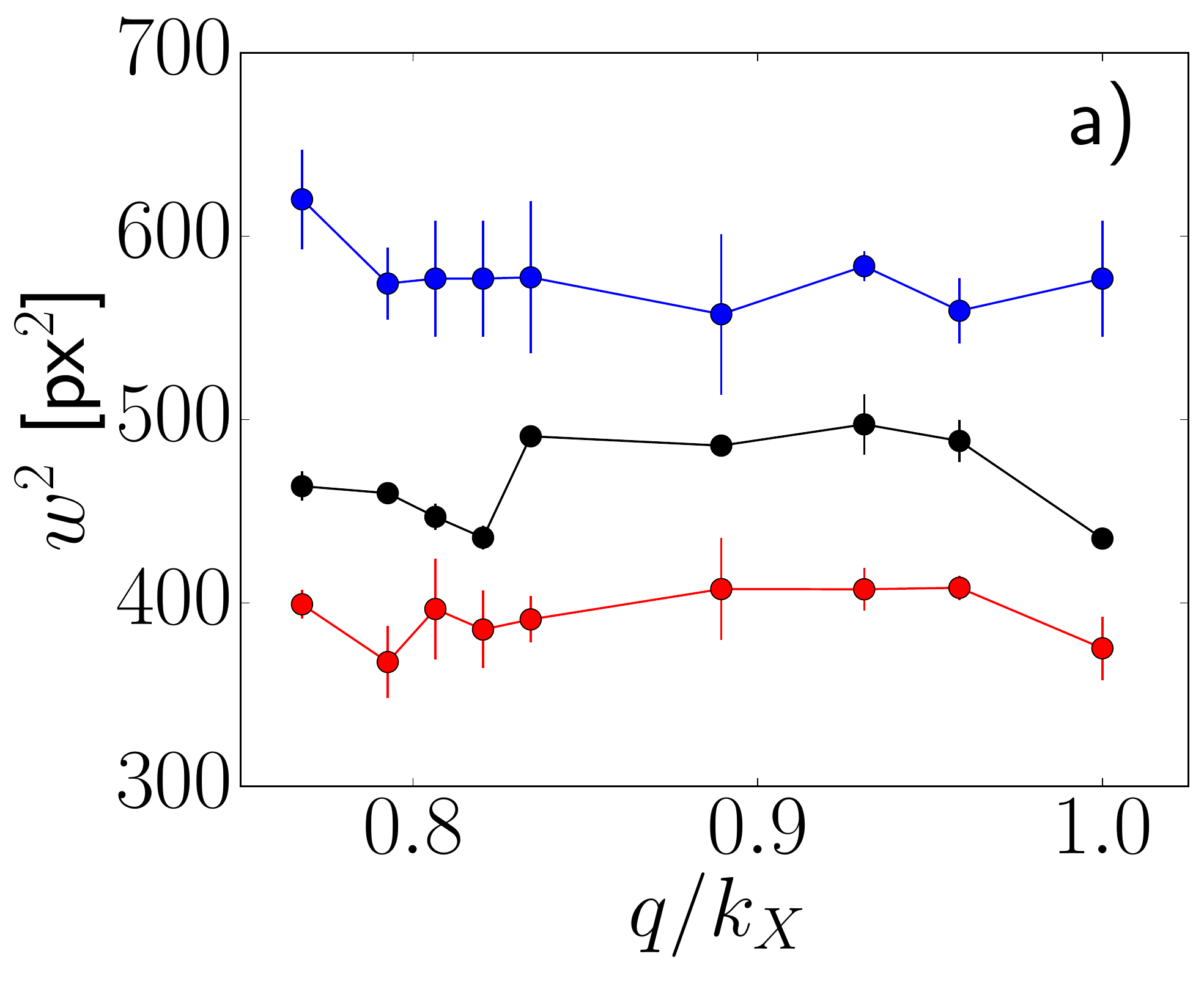}
    \includegraphics[width=0.492\linewidth]{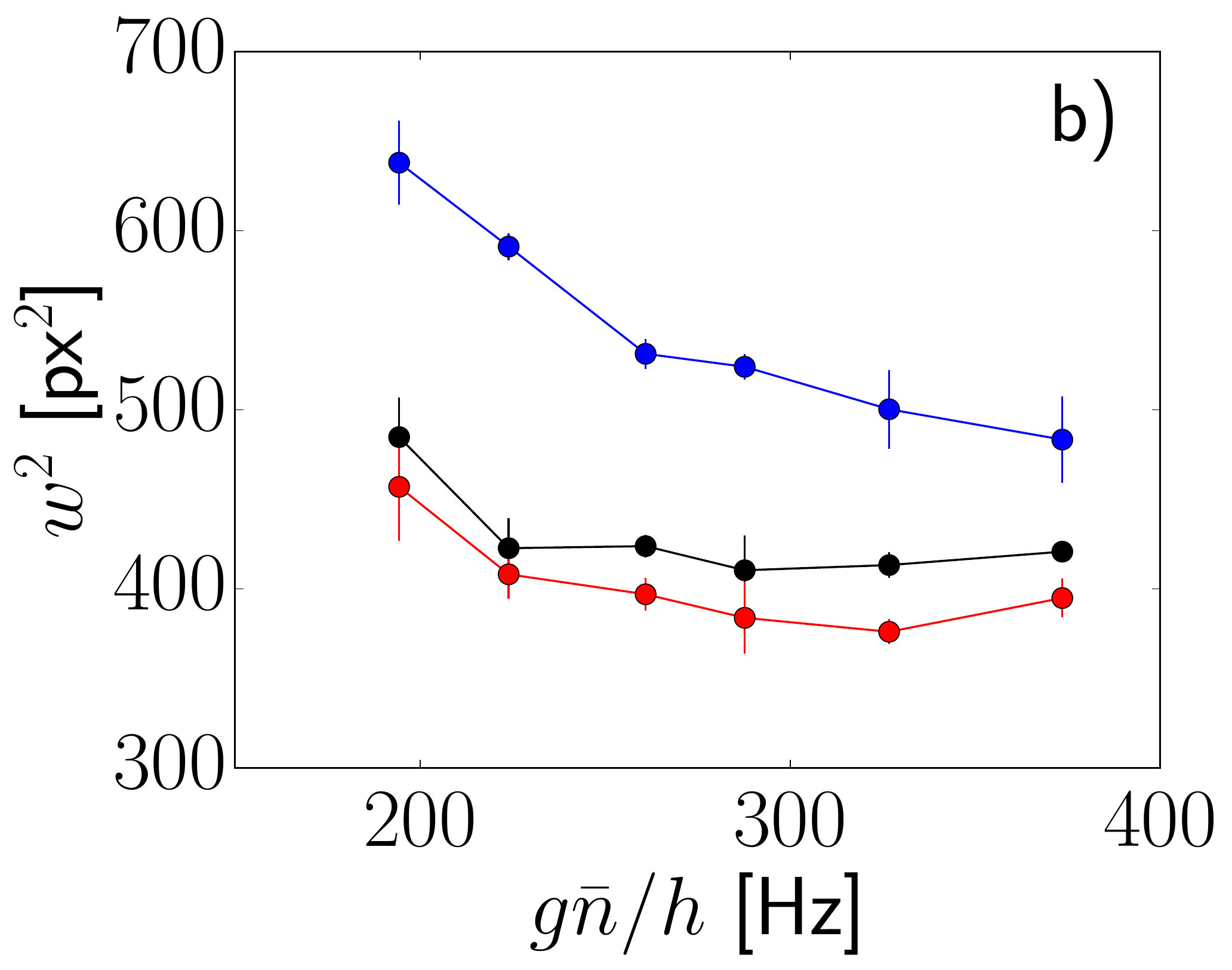}
  \caption{\label{fig:ntemp} Mean squared width, $w^2$, after decay from the Ground (red), Loop (black) and Excited (blue) prepared states, a) as a function of $q$ at an offset of 700~Hz. (Compare with Fig.~\ref{fig:qscans}a)), and b) as a function of interaction energy $g\bar{n}/h$ for $\Delta/h$=1.1~kHz. Error bars indicate the standard deviation from different measurements over all states. 
}
\end{figure}

An additional consequence of the nonlinear band structure is that the total energy of the looped band state should be more than the ground band state at the same $q$. We investigate the energy released from the different initial states by measuring the cloud width after decay. For each preparation sequence, we calculate the mean square width of the quasimomentum distribution, $w^2=\langle r^2 n(r) \rangle /\langle n(r)\rangle$, at times much longer than the decay time, where $r$ is the distance measured from the center of the BZ. As shown in Fig.~\ref{fig:ntemp}, the loop prepared states have larger final energy than the ground states, despite the fact they have identical lattice configurations and the loop state decays more slowly. The discrepancy in released energy is reduced, but does not vanish, when the decay rate gap closes, perhaps indicating either additional energy due to imperfect state preparation during the L sequence or a region of multi-valued bandstructure in which the loop and ground bands have the same stability. We note that the decay rates and energy released for the three different state preparations trend in opposite directions: the faster decaying G states have smaller final $w^2$, while the slower L and E states have larger final $w^2$. This supports a description of decay driven by dynamic instabilities, rather than energetic considerations.

The observed convergence of the decay rate of the G and L prepared states occurs at values of $q$, $\Delta$, and $g\bar{n}$ that agree qualitatively with the trends we expect from a mean-field calculation. However these points do not agree with a quantitative analysis of the region where a stable loop is predicted given the Bogliubov spectrum either in full 2D or restricted to modes in the direction of the acceleration in quasimomentum~\cite{supplement}. This discrepancy suggests the measured decay rates are due to a combination of effects beyond just the dynamic stability of the mean field state, including inhomogeneities across the lattice, state preparation infidelities and beyond mean-field correlations. It is clear, however, that the nature of the mean field state for a given preparation plays an important role in the relaxation. The decay mechanisms and their interplay with the transverse degrees of freedom is an interesting topic that could be further studied by adding a weak lattice in the third spatial dimension to control the magnitude of correlations and dispersion in that direction.

The authors thank M. Foss-Feig for helpful discussions. This work was partially supported by the ARO's Atomtronics MURI. RMW acknowledges partial support from the National Science Foundation under Grant No. PHYS-1516421.

\bibliography{library}

\section{Supplement}
\subsection{Lattice and state preparation}

All experiments begin with a $^{87}$Rb BEC with no discernible thermal fraction in the $| F = 1, m_F = -1\rangle$ internal hyperfine state, optically trapped with trap frequencies $(\nu_x, \nu_y, \nu_z) = (12(2), 40(4), 100(9))$~Hz. Control of the atom number, independent of trap parameters, is achieved by microwave removal of a fraction of atoms before the final stage of cooling. The lattice depth and offset $\Delta$ are determined from an experimentally calibrated model of the 2D lattice potential~\cite{PhysRevA.73.033605}. We parameterize the interaction energy by $g \bar{n}$, where $\bar{n}$ is the peak density averaged over a unit cell, $\bar{n} =  (2/\lambda^2) n_{1D}(z=0)$ and $n_{1D}$ is the 1D density along the lattice-free central tube. The interaction energy, $g \bar{n} (\lambda^2/2) \int d^2 r \left| \phi({\bf r})\right|^4$, depends on the size of the compressed localized Wannier function $\phi({\bf r})$, ($\phi({\bf r})$ is normalized to 1). For the lattice depths considered here, $(\lambda^2/2) \int d^2 r \left| \phi({\bf r})\right|^4 \simeq 4$, giving rise to an effective factor of four increase in interaction compared to a lattice-free case with the same average density. The value of $g \bar{n}$ is calculated either from an effective Thomas-Fermi approximation using the measured total atom number $N$ and trap frequencies (including the lattice), or from a full 3D ground state solution of the GP equation in the lattice. The two methods agree to $5~\%$.

The preparation sequences were empirically chosen to optimize the coherence of the final state BEC, while avoiding band excitations and minimizing dynamical instability decay during preparation. The BEC was initially loaded into a 8.9~E$_R$ lattice with positive offset $\Delta_1>0$ by turning on the lattice beams during 200~ms, followed by a 400~ms hold time. ($\Delta_1/h$ = 2.8~kHz for the E and L sequence and $\Delta_1/h$ = 1.7~kHz for the G sequence.) In order to minimize excitations during subsequent lattice manipulations, the lattice depth is then ramped in 0.5~ms to a depth of 10.6~E$_R$ and a larger offset $\Delta_2$. ($\Delta_2/h $ = 3.3~kHz for the E and L sequences, and $\Delta_2$ = 2.0~kHz for the G sequence.) 

After the increase in lattice depth, a magnetic field gradient is turned on to accelerate the BEC from $q=0$ to the final value near $q=k_X$ in a time $t_F \simeq 1$~ms. The acceleration time was chosen to prevent excitations, but to minimize dynamical decay during the acceleration. The switch from positive to negative offset in the E and L sequences, as well as the switch to the final offset value, only couple states with the same $q$, and were chosen to project the BEC onto particular states in the resulting band structure without residual band excitation. The switch from positive to negative offset was tested at $q=0$ by switching back and looking for excitation in the second BZ. No discernible excited fraction was observed for the lattice depth used. The GP simulations confirm that this combination of adiabatic and diabatic manipulations results in preparation of states with large overlap with the desired final states.

\subsection{ Image analysis} 
 The data was taken by absorption imaging the atom cloud after 21~ms time-of-flight, effectively measuring the momentum distribution. The lattice turn-off time of 1.5~ms was chosen to  ``band map'' the quasimomentum distribution $n(q)$ onto the free particle momentum distribution $n(p)$. (Near the band edges, it is impossible to be fully adiabatic, but this does not mix $q$ and only results in some mixing of $n(q)$ at points near the band edges that differ by reciprocal lattice vectors.) For the G and L sequences, some additional coherence decay occurs during the band mapping. The resulting momentum distribution is then divided into several regions $S$, $B_1$, $B_2$ and $K$. The relative position of $B_1$, $B_2$ and $K$ are fixed, but the position of region $S$ depends on $q$, since the signal in $S$ measures the initially prepared BEC state.  The signals in regions $B_1$ and $B_2$ measure the population outside of $S$ in the first and second BZ, respectively. These regions are chosen slightly smaller than the BZ to avoid ambiguous population mixed between different zones. The large triangles $K$ in the corners of the image are used to determine the background per pixel, $K /N_K$, assuming it is constant over the image and $N_K$ is the number of background pixels. We calculate the background-subtracted signal for the decay curves:
\begin{equation*}
\end{equation*}
where $N_X$ is the number of pixels in zone $X$ and $C$ is a single constant for each preparation $G$, $E$ or $L$, chosen so that the visibility decays to zero. 

The mean squared width, $w^2=\langle r^2 n(r) \rangle /\langle n(r)\rangle$, is calculated from images taken after the atom distribution has relaxed. We note that the band mapping technique modifies the initial energy distribution, and the observed time of flight $w^2$ does not directly represent the energy distribution in the lattice. Within a band, the measured $w^2$ is monotonically related to the energy in the band.

\subsection{Mean-field loop and stability regions}
\begin{figure}
\centering
  \includegraphics[width=0.98\linewidth]{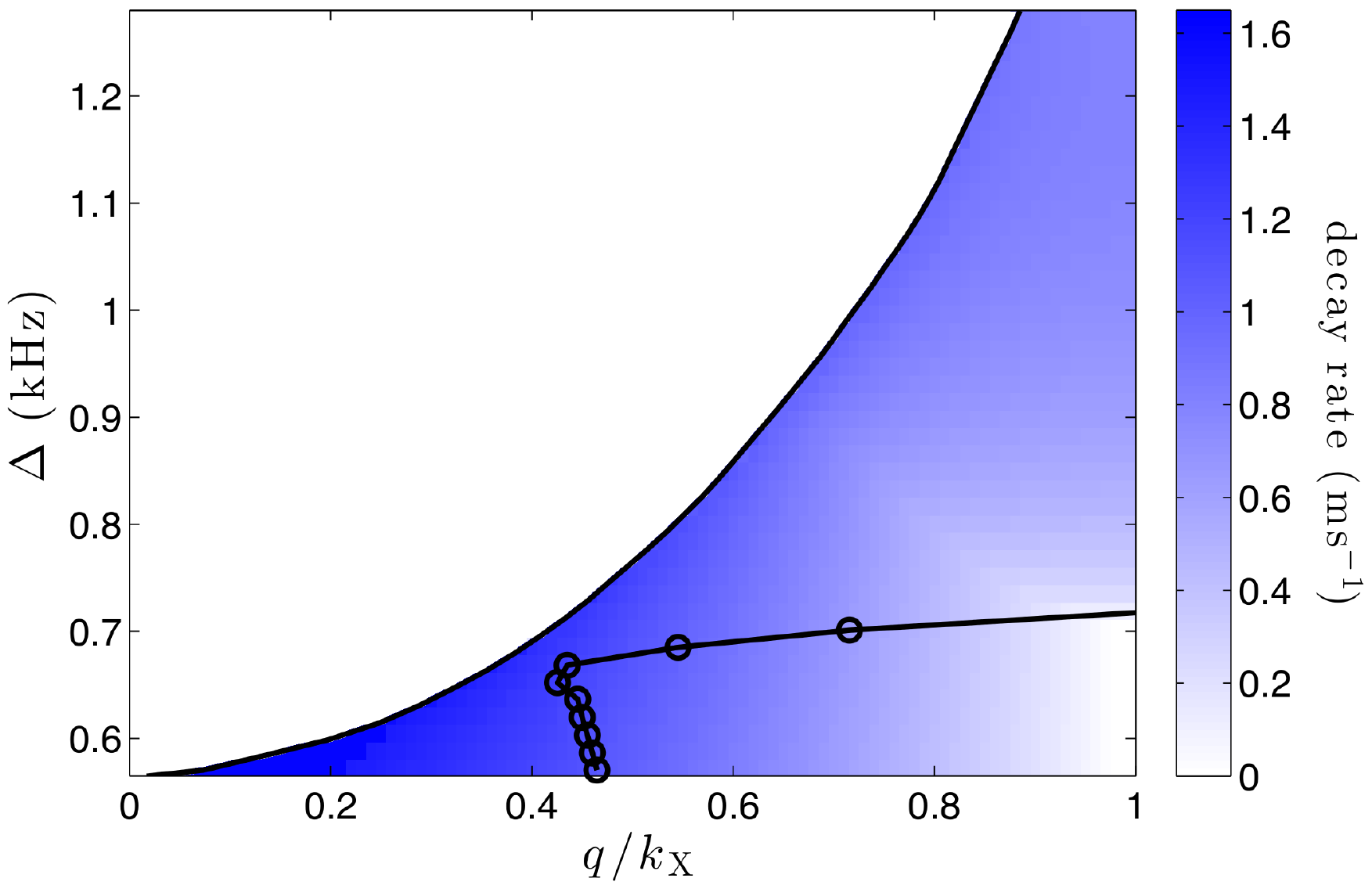}
	\caption{\label{fig:stabS} Presence and stability of the loop band for interaction energy $g\bar{n}/h=0.3~\rm{kHz}$ as a function of $\Delta$ and $q$. Multi-valued band structure is present in the blue region to the right of the black line. The color represents the decay rate for the condensate in the loop band, predicted by the Bogoliubov spectrum. Black points enclose a region where modes propagating in the direction of the acceleration in quasimomentum are stable, important because dynamic instabilities are more troublesome if seeded. }
\end{figure}

We model the experimental state preparation, the nonlinear band structure, and the stability of the states with an effective two-dimensional mean-field theory described by the energy functional
\begin{align}
\label{GPenergy}
E = \int d^2 r\, \psi^*(\mathbf{r}) \left( \frac{-\hbar^2}{2m} \nabla^2 + V_\mathrm{lat}(\mathbf{r}) + \frac{g_\mathrm{2D}}{2} | \psi(\mathbf{r}) |^2 \right) \psi(\mathbf{r})
\end{align}
and the corresponding time-dependent Gross-Pitaevskii equation
\begin{align}
\label{GPE}
i \hbar \partial_t \psi(\mathbf{r}) = \left( \frac{-\hbar^2}{2m} \nabla^2 + V_\mathrm{lat}(\mathbf{r}) + g_\mathrm{2D} | \psi(\mathbf{r}) |^2 \right) \psi(\mathbf{r}),
\end{align}
where  $V_\mathrm{lat}(\mathbf{r})$ is the lattice potential
\begin{align}
V_\mathrm{lat} (\mathbf{r}) &= \frac{\Delta}{4} \left( \cos (2 k_R x) - \cos(2 k_R y) \right)  \nonumber \\
&+   V_0 \cos (2 k_R x) \cos (2 k_R y),
\end{align}
$V_0$ is the lattice depth,  and $g_\mathrm{2D} = g \bar{n} \lambda^2 / 2$ is the effective interaction coupling.  In our dynamical simulations of Eq.~(\ref{GPE}) to model the state preparation, we consider unit cells of area $\lambda^2/2$ and impose periodic boundary conditions in the phase-gradient of $\psi(\mathbf{r})$.  
We calculate the nonlinear band structure following the method of~\cite{PhysRevA.86.063636}.  We find the stationary Bloch solutions of Eq.~(\ref{GPE}), which have the form $\psi_{n \mathbf{q}}(\mathbf{r}) = e^{i \mathbf{q} \cdot \mathbf{r}} u_{n \mathbf{q}} (\mathbf{r})$ where $n$ is the band index and $\mathbf{q}$ is the quasimomentum in two dimensions.  We work in the reciprocal space, and expand $u_{n \mathbf{q}}(\mathbf{r}) = \sum_{\mathbf{k}} c_{n \mathbf{k}} e^{i \mathbf{k} \cdot \mathbf{r}} $.  We find $ \{ c_{n \mathbf{k} } \}$ numerically for all relevant values of $n$ and $\mathbf{q}$, reconstruct $\psi_{n \mathbf{q}}$, then find the energy of the state using Eq.~(\ref{GPenergy}).      

The expected parameter regime for looped band structure at a representative interaction strength of $g\bar{n}=0.3$~kHz is shown in Fig.~\ref{fig:stabS}.  Here, the blue region indicates the presence of looped band structure, and the shading corresponds to the decay rate of the loop state at a given quasimomentum $\mathbf{q} = (q,0)$.  We model decay rates using a linear stability analysis, which in this mean-field framework corresponds to a Bogoliubov treatment.  We derive the Bogoliubov equations by substituting $\psi_{n \mathbf{q} }(\mathbf{r}) \rightarrow  \psi_{n \mathbf{q} }(\mathbf{r})  + \delta \sum_\mathbf{p} \left( \mathcal{U}_{n , \mathbf{q}+\mathbf{p}} e^{i (\mathbf{p} \cdot \mathbf{r} - \omega t) } +  \mathcal{V}^*_{n , \mathbf{q} - \mathbf{p}} e^{-i (\mathbf{p} \cdot \mathbf{r} - i \omega t )} \right)  $ in Eq.~(\ref{GPE}) and linearizing in $\delta$.  We diagonalize these equations to obtain the Bogoliubov spectrum.  The imaginary part of this spectrum corresponds to decay of the equilibrium state at the rate $\mathrm{Im}[\omega]$.  In Fig.~\ref{fig:stabS}, the blue shading represents the maximum value of this rate for all values of $\mathbf{p} = (p_x,p_y)$, while the black circles enclose a region where modes propagating in the direction of the quasimomentum acceleration are dynamically stable, and correspondingly exhibit no decay.

\end{document}